\documentclass[%
reprint,
superscriptaddress,
showpacs,
nofootinbib,
nobibnotes,
 amsmath,amssymb,
 aps,
prc,url
floatfix,
]{revtex4-1}

\usepackage{graphicx}% Include figure files
\usepackage{dcolumn}% Align table columns on decimal point
\newcolumntype{.}{D{.}{.}{-1}}
\usepackage{bm,lineno}% bold math
\modulolinenumbers[5]
\usepackage{float,array,booktabs,amsmath,amssymb}
\begin{document}

\preprint{APS/123-QED}

\title{\boldmath Proton capture on $^{30}$P in novae: On the existence of states at $6.40$~MeV and $6.65$~MeV in $^{31}$S
}
\author{M.~Kamil}
\affiliation{Department of Physics and Astronomy, University of the Western Cape, P/B X17, Bellville 7535, South Africa}%
\author{S.~Triambak}
\email{striambak@uwc.ac.za}
\affiliation{Department of Physics and Astronomy, University of the Western Cape, P/B X17, Bellville 7535, South Africa}%
\author{G.\,C.~Ball}
\affiliation{TRIUMF, 4004 Wesbrook Mall, Vancouver, British Columbia V6T 2A3, Canada.}
\author{V.~Bildstein}
\affiliation{Department of Physics, University of Guelph, Guelph, Ontario N1G 2W1, Canada}%
\author{A.~Diaz Varela}
\affiliation{Department of Physics, University of Guelph, Guelph, Ontario N1G 2W1, Canada}%
\author{T.~Faestermann }
\affiliation{Physik Department, Technische Universit\"{a}t M\"{u}nchen, D-85748 Garching, Germany}%
\author{P.\,E.~Garrett}
\affiliation{Department of Physics, University of Guelph, Guelph, Ontario N1G 2W1, Canada}%
\affiliation{Department of Physics and Astronomy, University of the Western Cape, P/B X17, Bellville 7535, South Africa}%
\author{F.\,Ghazi Moradi}
\affiliation{Department of Physics, University of Guelph, Guelph, Ontario N1G 2W1, Canada}%
\author{R.~Hertenberger}
\affiliation{Fakult\"{a}t f\"{u}r Physik, Ludwig-Maximilians-Universit\"{a}t M\"{u}nchen, D-85748 Garching, Germany}%
\author{N.\,Y.~Kheswa}
\affiliation{iThemba LABS, P.O. Box 722, Somerset West 7129, South Africa}%
\author{N.\,J.~Mukwevho}
\affiliation{Department of Physics and Astronomy, University of the Western Cape, P/B X17, Bellville 7535, South Africa}%
\author{B.\,M.~Rebeiro}
\altaffiliation[Present address: ]{Department of Physics, McGill University, Montr\'eal, Qu\'ebec, Canada H3A 2T8.}
\affiliation{Department of Physics and Astronomy, University of the Western Cape, P/B X17, Bellville 7535, South Africa}%
\author{H.\,-F.~Wirth}
\affiliation{Fakult\"{a}t f\"{u}r Physik, Ludwig-Maximilians-Universit\"{a}t M\"{u}nchen, D-85748 Garching, Germany}%
%  \email{Second.Author@institution.edu}
% \affiliation{%
%  Authors' institution and/or address\\
%  This line break forced with \textbackslash\textbackslash
% }%
% 
% \collaboration{MUSO Collaboration}%\noaffiliation
%  \homepage{http://www.Second.institution.edu/~Charlie.Author}
% \affiliation{
%  Second institution and/or address\\
%  This line break forced% with \\
% }%
% \affiliation{
%  Third institution, the second for Charlie Author
% \affiliation{%
%  Authors' institution and/or address\\
%  This line break forced with \textbackslash\textbackslash
% }%
% 
% % 
% % \author{V.~Pesudo}%
% % %  \affiliation{University of the Western Cape, P/B X17, Bellville 7535, South Africa}
% %   \affiliation{iThemba LABS, P.O. Box 722, Somerset West 7129, South Africa}
% %  \author{B.~Singh}
% %  \affiliation{University of the Western Cape, P/B X17, Bellville 7535, South Africa}
 
\date{\today}% It is always \today, today,
 \begin{abstract}
We use a high resolution $^{32}{\rm S}(d,t)$ measurement to investigate the claimed existence of a 6401(3)~keV state in $^{31}$S that may affect the $^{30}{\rm P}(p,\gamma)$ nuclear reaction rate in oxygen-neon (ONe) novae. Our  data are shown to exclude the null hypothesis - that the state does not exist - with high significance. Additionally, the data  also suggest the existence of a hitherto unreported state at 6648(4)~keV.  This state corresponds to a $^{30}{\rm P}(p,\gamma)$ resonance at $517(4)$~keV, located below the higher edge of the Gamow window for peak nova temperatures of about 0.4~GK. 

%This observation supports the existence of the state. 
\end{abstract}

%\pacs{ }% PACS, the Physics and Astronomy
\maketitle
\noindent
%\linenumbers
\section{Introduction}
 A current topic of interest in nuclear astrophysics is  nucleosynthesis in novae~\cite{jose:01,jose:07}. The latter are transient explosive phenomena that occur in binary star systems, which comprise white dwarfs and their companion main-sequence stars. 
 In this context, observational data from space and terrestrial based telescopes~\cite{Hayward,mnras:87}, isotopic analyses of presolar meteoritic grains~\cite{Amari:01} and hydrodynamic modeling of nova explosions~\cite{Jose:99, Iliadis:02} have played a critical role towards a better understanding of classical novae. An important ingredient in one class of nova models (ONe novae) is the $^{30}{\rm P}(p,\gamma)^{31}{\rm S}$ nuclear reaction rate~\cite{jose:01}. This reaction rate significantly impacts nucleosynthesis in the Si-Ca mass region~\cite{jose:01} and has a large uncertainty, mainly because of the present unavailability of intense $^{30}$P beams for both direct and indirect measurements. Nevertheless, despite this limitation, there has been considerable progress in constraining the $^{30}{\rm P}(p,\gamma)^{31}{\rm S}$ reaction rate 
 using indirect probes~\cite{Budner,Bennett:16,Ma:07,Wrede:07,Wrede:09,Parikh:11, Irvine,Anu1,Bennett:18,Anu2,Doherty1,Doherty2,Setoodehnia,Jenkins}. 
 
 Resonances in the $^{30}{\rm P}(p,\gamma)$ reaction compete with $^{30}$P $\beta^+$ decay ($t_{1/2} \approx 2.5$~min) at peak nova temperatures of around $0.1$--$0.4$~GK, thereby controlling the nucleosynthesis path towards heavier species~\cite{jose:01}. The dominant resonances are expected to be in the range of $\sim$~600~keV~\cite{JoseBook} above the 6131~keV proton emission threshold in $^{31}$S. This corresponds to an excitation region $6.1 \lesssim E_x \lesssim 6.7$~MeV in $^{31}$S. Consequently, there have been several experimental investigations~\cite{Budner,Bennett:16,Ma:07,Wrede:07,Wrede:09,Parikh:11,Irvine,Anu1,Bennett:18,Anu2,Doherty1,Doherty2,Setoodehnia,Jenkins} of $^{31}$S excited states in this energy range, to indirectly determine $^{30}{\rm P}(p,\gamma)$ resonance strengths to these states.
 Available spectroscopic information from these experiments have led to a debate about the existence of a possible state at 6.4~MeV, which may significantly contribute to the total $^{30}{\rm P}(p,\gamma)$ reaction rate.  

 A 6400(3)~keV state  was first reported by Wrede~\textit{et al.}~\cite{Wrede:07}, who used a $^{31}{\rm P}(^3{\rm He},t)$ experiment. It was identified as a $d$-wave ($\ell = 2$) resonance and shown to dominate the reaction rate in the temperature range of interest. Subsequent $^{32}{\rm S}(d,t)$ work at the same facility reported the state at $E_x = 6398(6)$~keV~\cite{Wrede:09}. Shortly after, Parikh~\textit{et al.}~\cite{Parikh:11} used a higher resolution $^{31}{\rm P}(^3{\rm He},t)$ measurement to report the same level at 6403(2)~keV. This was followed by a high resolution $^{32}{\rm S}(d,t)$ experiment~\cite{Irvine} that confirmed its existence at 6402(2)~keV. 
 %The authors of Ref.~\cite{Irvine} went one step further and used a comparison with known levels in the mirror $^{31}$P nucleus to assign the state a spin value of $J = 7/2$. 
 Unfortunately, the above claims were not adequately supported by a rigorous description of the statistical analyses performed on their data sets. Nevertheless, the state is now listed at $E_x = 6401(3)$~keV on the Evaluated Nuclear Structure Data File (ENSDF)~\cite{ensdf} at the National Nuclear Data Center (NNDC). This corresponds to a $^{30}{\rm P}(p,\gamma)$  resonance at $E_r = 270(3)$~keV. 
 
Contrary to the above, extensive $\gamma$-ray spectroscopy experiments with the  GAMMASPHERE and the GRETINA arrays, following the $^{28}{\rm Si}(\alpha,n)$~\cite{Doherty1,Doherty2}, $^{12}{\rm C}(^{20}{\rm Ne},n)$~\cite{Jenkins} and the $^{30}{\rm P}(d,n)$~\cite{Anu2} reactions\footnote{The $^{30}{\rm P}(d,n)$ experiment was performed in inverse kinematics.} did not show any evidence of $\gamma$-transitions from the claimed 6.4~MeV state.
More recently, $\gamma$-ray spectroscopy following $^{31}{\rm Cl}$~$\to$~$^{31}$S ${\beta^+}$ decay~\cite{Bennett:16,Bennett:18} and independent $^{32}{\rm S}(p,d)$ studies~\cite{Setoodehnia,tamu} also did not confirm the existence  of this level.
%\footnote{The $^{32}{\rm S}(p,d)$ data~\cite{Setoodehnia} were limited by experimental resolution.} 

Motivated by these inconsistent observations and the lack of sufficient detail provided for the statistical analysis in the transfer/charge-exchange work~\cite{Wrede:07,Wrede:09,Parikh:11,Irvine}, we investigated the relevant excitation energy region in $^{31}$S with a high resolution $^{32}{\rm S}(d,t)$ measurement. %Experimental details and results are presented below.   sxs

\section{Apparatus}
%\section{Experimental details}
The experiment was performed at the Maier-Leibnitz-Laboratorium (MLL) tandem accelerator facility in Garching, Germany. Approximately 400~nA of 23~MeV deuterons were incident on an $\sim 120~\mu$g/cm$^2$-thick natural ZnS target, that was evaporated on an $\sim$ 20~$\mu$g/cm$^2$ natural carbon backing. The $^{32}{\rm S}+d$ reaction products were momentum analyzed with the Q3D magnetic spectrograph~\cite{Loffler:73,Thomas}, whose solid angle acceptance was kept fixed at $\sim 14.6$~msr throughout the experiment. The focal plane detector for the spectrograph consisted of two gas proportional counters (with $\sim$~500 mbar of isobutane gas) and a 7-mm-thick plastic scintillator~\cite{Thomas}. The energy losses registered in the proportional counters and the residual energy deposited in the plastic scintillator were used to discriminate the tritons from other ejectiles, while a cathode strip foil in the second proportional counter provided high-resolution position information for the tritons.  Since a previous $^{32}{\rm S}(d,t)$ experiment  at the same facility (Irvine~{\it et al.}~\cite{Irvine}) claimed a $> 4\sigma$ signal for the 6.4~MeV state at $\theta_{\rm lab} = 53.75^\circ$, we acquired data at $\theta_{\rm lab} = 55^\circ$. For comparison, additional data were also acquired at $\theta_{\rm lab} = 45^\circ$, albeit with significantly lower statistics. 

\section{Data analysis}
\begin{figure}[t]
\includegraphics[scale=0.34]{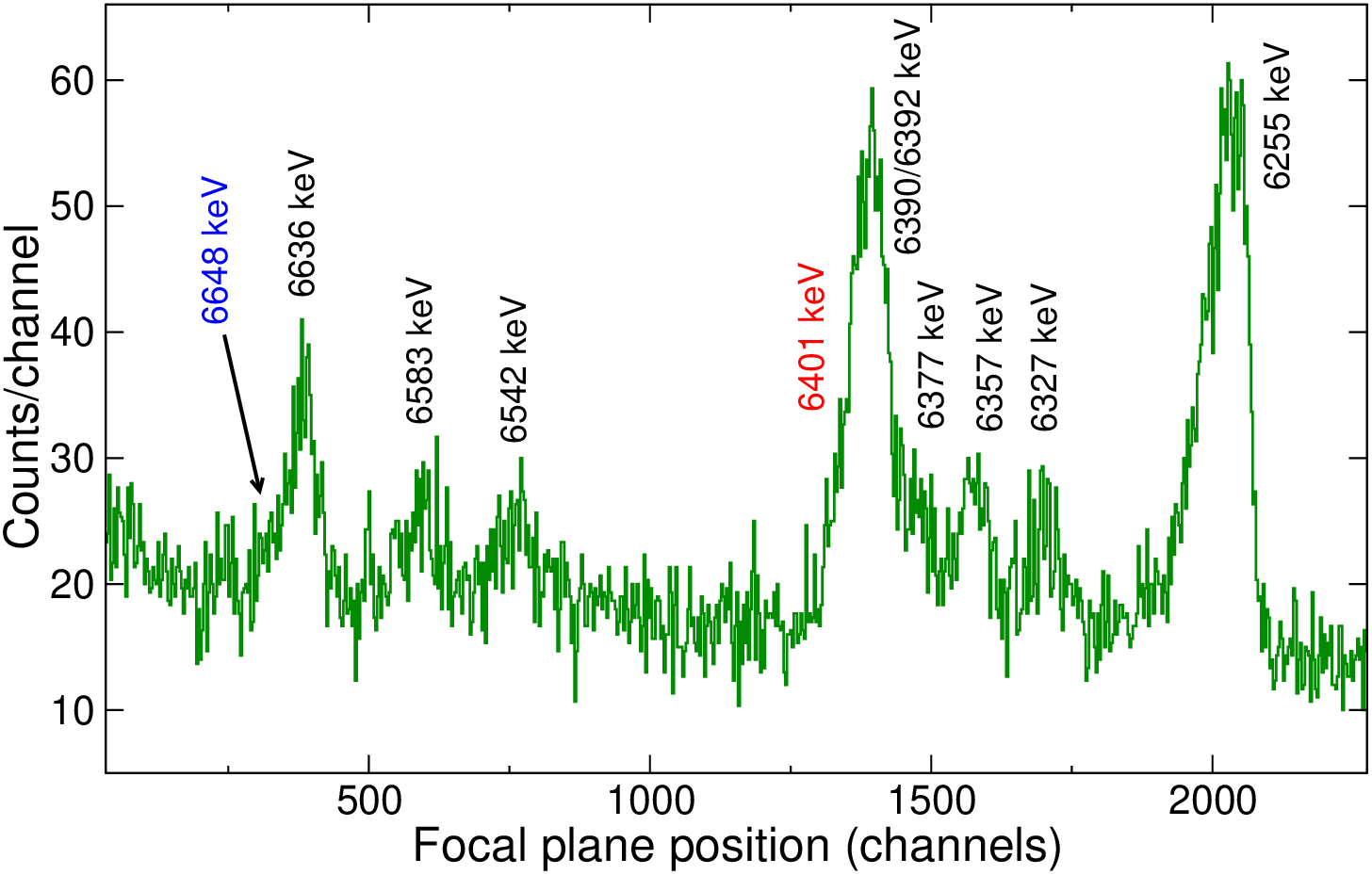}% Here is how to import EPS art
\caption{\label{fig:spectrum}$^{32}{\rm S}(d,t)$ spectrum at $\theta_{\rm lab} = 55^\circ$. This histogrammed spectrum was rebinned by a factor of 3, only for visualization purposes. 
%The rebinning procedure was performed such that for a rebinning factor $k$, $\displaystyle x_j = \frac{1}{k} \sum_{k(j-1)+1}^{kj} x_i$.  . 
The purported 6401~keV state in $^{31}$S is labeled in red. The 6648~keV state is a new level reported in this work. Nominal values for the other states are taken from Ref.~\cite{ensdf}.
}
\end{figure}

Figure~\ref{fig:spectrum} shows the $\theta_{\rm lab} = 55^\circ$ triton spectrum
%\footnote{The rebinning procedure for this histogram was performed such that, for a rebinning factor $k$, $\displaystyle x_j  = \frac{1}{k} \sum_{i_\mathrm{ini}}^{kj} x_i$ and  $\displaystyle y_j  = \frac{1}{k} \sum_{i_\mathrm{ini}}^{kj} y_i$, where $i_\mathrm{ini} =  k(j-1)+1$.} 
from this experiment, with statistics comparable to those reported by Irvine~{\it et al.}~\cite{Irvine}. The triton peaks in our data had full widths at half maximum (FWHM) of $\sim$12-15~keV, their resolution being mainly limited by the target thickness. 
%\footnote{Our resolution is poor compared to Irvine~{\it et al.}~\cite{Irvine} mainly due to the thickness of our target.} 
Unlike Fig.~\ref{fig:spectrum}, the histograms used for data analysis were not rebinned  to avoid spurious binning artefacts. These had the original binning provided by the data acquisition system, with each bin corresponding to 0.35~mm in the focal plane. The spectra were analyzed assuming a model 
\begin{equation}
y_i(\bm{\theta}) = A \int_{{\rm bin}~i}F(x_i;\mu,\sigma,l,\xi) dx + B, 
\label{eq:model}
\end{equation}
for each peak. In the above, $F(x_i;\mu,\sigma,l,\xi) = \xi T(x) + (1-\xi) G(x)$, where $T(x)$ is the convolution of a Gaussian with a low (particle) energy exponential tail, $\xi$ being its relative strength with respect to the pure Gaussian, $G(x)$. The functional form of $T(x)$ is described in Refs.~\cite{Triambak1,Triambak2}. $\bm{\theta}$ contains the maximum likelihood estimators (MLEs)~\cite{CowanBook}, $A, \mu, \sigma$, $l$, $\xi$ and $B$. Here, $A$ represents the peak amplitude, $\mu$ is its centroid, $B$ is a flat background, the peak FWHM = $2 \sqrt{2 \ln 2}~\sigma$, and $l$ is the decay length of the exponential tail. The tail parameters, $\xi$ and $l$ mainly depend on the detector response function. The dispersion of the Q3D is known to change in a manner such that larger tail components are observed for higher energy particles, compared to particles with lower energy. This is evident in Fig.~\ref{fig:spectrum}, where the 6255~keV peak with the highest triton energy has a prominently visible low-energy tail compared to the others.    

In the first stage of analysis, the triton peaks were fit using a Levenberg-Marquardt algorithm~\cite{NR}, assuming that the entries in each bin ($n_i$) are independent and Poisson distributed. In such a scenario, one obtains  the MLEs via a minimization of the quantity~\cite{Baker,pdg},
\begin{equation}
-2 \ln \lambda(\bm{\theta}) = 2 \sum_{i=1}^{N~{\rm bins}}\left[y_i(\bm{\theta}) - n_i + n_i \ln \frac{n_i}{y_i(\bm{\theta})}\right],
\label{eq:min}
\end{equation}
where $\lambda(\bm{\theta})$ is the likelihood ratio defined in Refs.~\cite{Baker,pdg}. From this analysis we obtained the $\xi$ and $l$ MLEs for the 6255~keV peak to be about $0.9$ and $64$~channels, for the $55^\circ$ data set.\footnote{Similar values were obtained for the $\theta_\mathrm{lab} = 45^\circ$ data.}  However, given the 
modest statistics in our spectra (c.f. Fig.~\ref{fig:spectrum}), we could not fit the other peaks similarly, with the tail parameters kept free. Consequently, we used other Q3D data~\cite{Rebeiro1,Rebeiro2,Rebeiro3,Jespere,Kamil} to determine $\xi$ and $l$ for peaks in the 6400~keV region of interest (ROI). These independent data sets~\cite{Rebeiro1,Rebeiro2,Rebeiro3,Jespere,Kamil} showed that peaks located at about the same focal plane position required $\xi \approx 0.3 $ and $l \approx 9.0$ channels to correctly describe their lineshapes. 
This information allowed a realistic response function to be incorporated in our analysis. The initial fits to the peaks in the $55^\circ$ spectrum were performed with the FWHM as unconstrained free parameters. This preliminary analysis provided critical guidance to proceed to the next stage.
For the null hypothesis $(H_0)$, we assume the absence of a state at $\approx$ 6401~keV.  Although the fits yielded reasonably good agreement with the data, the extracted FWHM  was
80(3)~channels for the 6390/6392 keV peak\footnote{In actuality there are three closely-spaced states reported at 6390~keV~\cite{Bennett:16,Bennett:18}, 6392~\cite{Doherty1,Doherty2} and 6394~\cite{ensdf}~keV. A recent $^{32}{\rm S}(p,d)$ measurement~\cite{Setoodehnia} did not show a strong population of the $11/2^+$ 6394~keV state. Since the other two states are only 2~keV apart and we are limited by experimental resolution, we assume a single peak in this energy region, with its centroid as a free parameter.}. This is unexpectedly large, and may be compared to FWHM values of 52(10) channels and 43(3) channels for the 6357 and 6255~keV states, respectively. As we describe in greater detail below, the nearly factor-of-two larger peak width at around channel 1400 (compared to channel 2000) is contrary to expectations, given the known variation in the Q3D dispersion with particle energy. These results already suggested the possibility of an additional peak in the 6400~keV ROI.

Considering the above, a more reasonable approach would be to assume that the lineshape parameters do not vary appreciably for the three triton peaks corresponding to the 6390/6392, 6377 and 6357~keV states (c.f. Fig.~\ref{fig:spectrum}).  Based on this premise we fitted the peaks corresponding to the 6390/6392 and 6377~keV states, with their FWHM fixed at 52 channels, the MLE value for the 6357~keV peak. Next, the maximum likelihood (ML) estimation procedure was used to perform a goodness-of-fit (GOF) test, with the test-statistic being $ t_{\bm{\hat{\theta}}} = -2\ln\lambda(\bm{\hat{\theta}})$~\cite{Baker}, where the $\bm{\hat{\theta}}$ represent ML values of the parameters. 
Here again the null hypothesis assumed only two peaks (corresponding to the 6390/6392 and 6377 keV states) in the ROI.
The significance of such a GOF test was obtained from its $p$-value, the probability of obtaining data with $t_{\bm{\theta}} \ge t_{\bm{\hat{\theta}}}$. Under certain conditions this can be evaluated by assuming Wilks' theorem~\cite{Wilks}, which states that $t_{\bm{\theta}}$ follows an asymptotic $\chi^2_\nu$ distribution, for $\nu$ degrees of freedom. Based on this, we obtain the null hypothesis $p$-value to be $1.4\times10^{-3}$ for the $55^\circ$ data. On the other hand, including an additional peak for the  6401~keV state (the alternative hypothesis, $H_1$) yielded $p = 0.63$. This clearly showed that under the assumptions mentioned above, the null hypothesis does not adequately describe the observed data.  

However, despite the seemingly plausible analysis described above, it can be critiqued that the GOF test relies heavily on the presence of a nearby high-statistics singlet peak. It is also evident that the 6357~keV peak had insufficient counts to determine its FWHM with reasonable precision, while being separated from the peak of interest by more than 100 channels. More importantly, Wilks' theorem does not hold in scenarios when parameters are at boundary points~\cite{Chernoff,Self} or when the model is expressed as a mixture of probability density functions (pdfs)~\cite{mixture}. Due to these considerations we reanalyzed the data using a more rigorous approach, described below. 

\begin{figure}[t]
\includegraphics[scale=0.34]{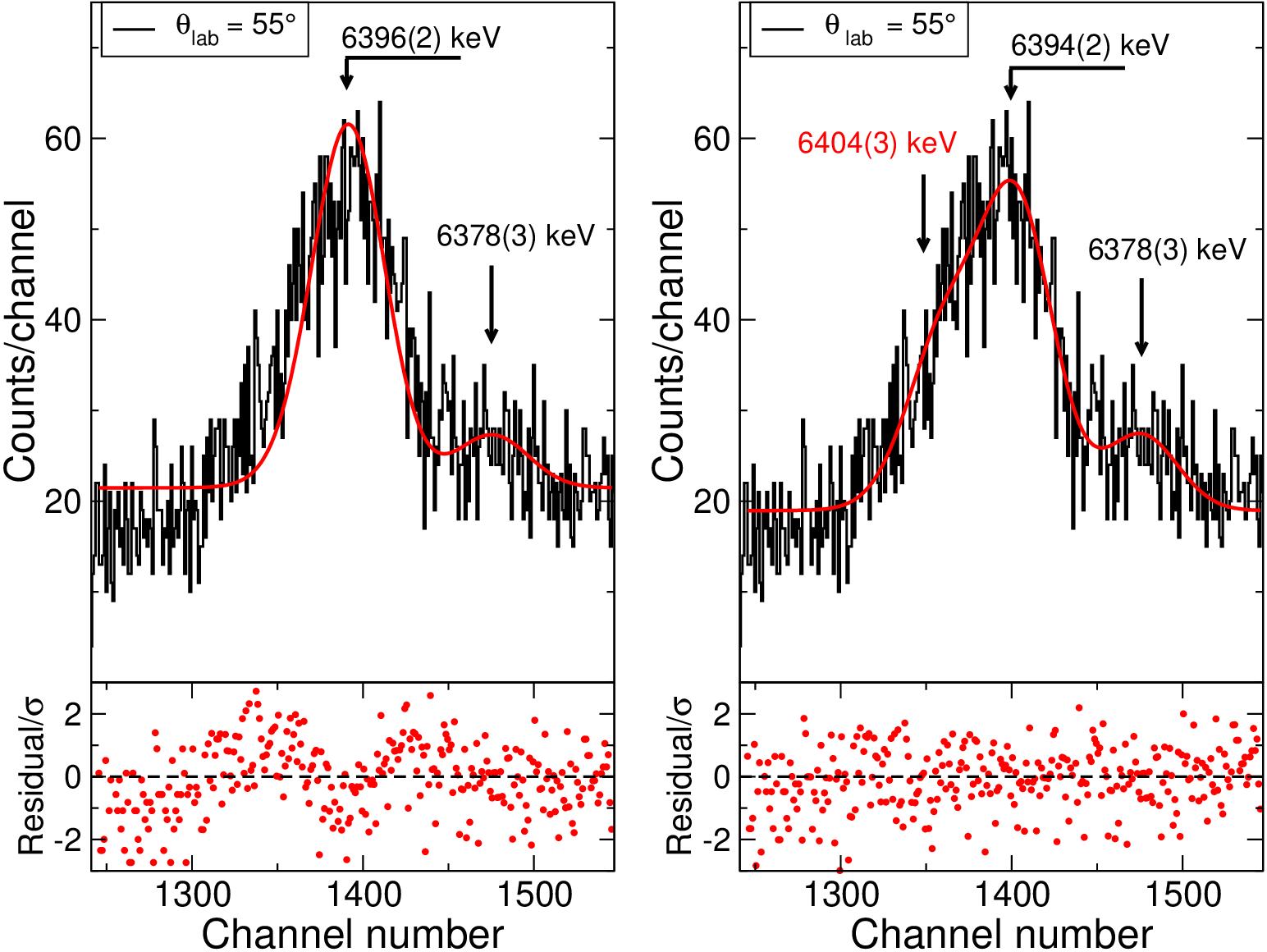}% Here is how to import EPS art
\caption{\label{fig:restricted}Left panel: Fit results for the $55^\circ$ data assuming the null hypothesis, which does not include a state at 6401~keV. The FWHMs of these peaks were constrained as described in the text. Right panel: A similar fit to the same data assuming that an additional state exists in the 6400~keV region. 
}
\end{figure}

The first step was to extract a more reliable estimate of the peak FWHM parameter for the ROI. For this we resort to the other Q3D data mentioned previously~\cite{Rebeiro1,Rebeiro2,Rebeiro3,Jespere, Kamil}, because of the dearth of high-statistics singlet peaks in our $^{32}{\rm S}(d,t)$ spectra. 
The FWHM values obtained from these independent data sets were recorded and fitted to a quadratic function to quantify the variation in peak widths with focal plane position. 
We find that in all data sets the peak FWHMs reduce with decreasing particle energy, by around 22-26\%, in the range from 2000 channels to 200 channels. This reduction is consistent with expectations~\cite{Loffler:73}. Once this information was determined, we extracted the expected FWHM for the 6390/6392~keV peak, using the measured width of the 6255~keV peak as a reference. For the $55^\circ$ data, we conservatively determine this expected FWHM to be 37(6) channels. Next we fitted the peaks in the ROI using an extended maximum likelihood procedure, placing approximately $\pm 95\%$ confidence level (CL) constraints on the FWHM. Assuming that the FWHM parameter is normally distributed, such a scenario required a modification of Eq.~\eqref{eq:min} to
\begin{equation}
\label{eq:min2}
%\begin{split}
-2 \ln \lambda(\bm{\theta})  = 2 \sum_{i=1}^{N~{\rm bins}}\left[y_i(\bm{\theta}) - n_i + n_i \ln \frac{n_i}{y_i(\bm{\theta})}\right] 
 + \left(\frac{f-F}{\sigma_F}\right)^2,
%\end{split}
\end{equation}
where $F$ is the expected value of the FWHM and $f$ is a variable within the interval $[F-2\sigma_F, F+2\sigma_F]$.  The fit results for the $55^\circ$ data are shown in the left panel of Fig.~\ref{fig:restricted}. 
%For the left panel (the null hypothesis fit), we note that FWHM parameter converges to the upper bound value of 49 channels.

% \footnote{The FWHM parameter also converges to its upper bound in the other two null hypothesis fits, shown in Figs.~\ref{fig:restricted2} and \ref{fig:fits3}. This indicates an excess of counts in the ROI under consideration for the three plots.} 
% %The extracted $t_{\bm{\hat{\theta}}}$ for both the null and the alternative hypotheses were then used to determine $p$-values.   
%
Due to the various pitfalls associated with the validity of Wilks' theorem, we next used toy Monte Carlo simulations to generate $10^5$ synthetic data sets for the GOF tests. These data were used to determine the distribution of $t_{\bm{\theta}}$, under the assumption that $H_0$ was the true model to describe the data. For this scenario, the GOF analysis yielded  
%are shown in Fig.~\ref{fig:null}, which 
$p < 10^{-5}$ for the $55^\circ$ data. This invalidates the null hypothesis with high significance. On repeating the analysis with an additional 6401~keV state peak within this ROI, we obtained a $p$-value of 0.54. 
\begin{figure}[t]
\includegraphics[scale=0.34]{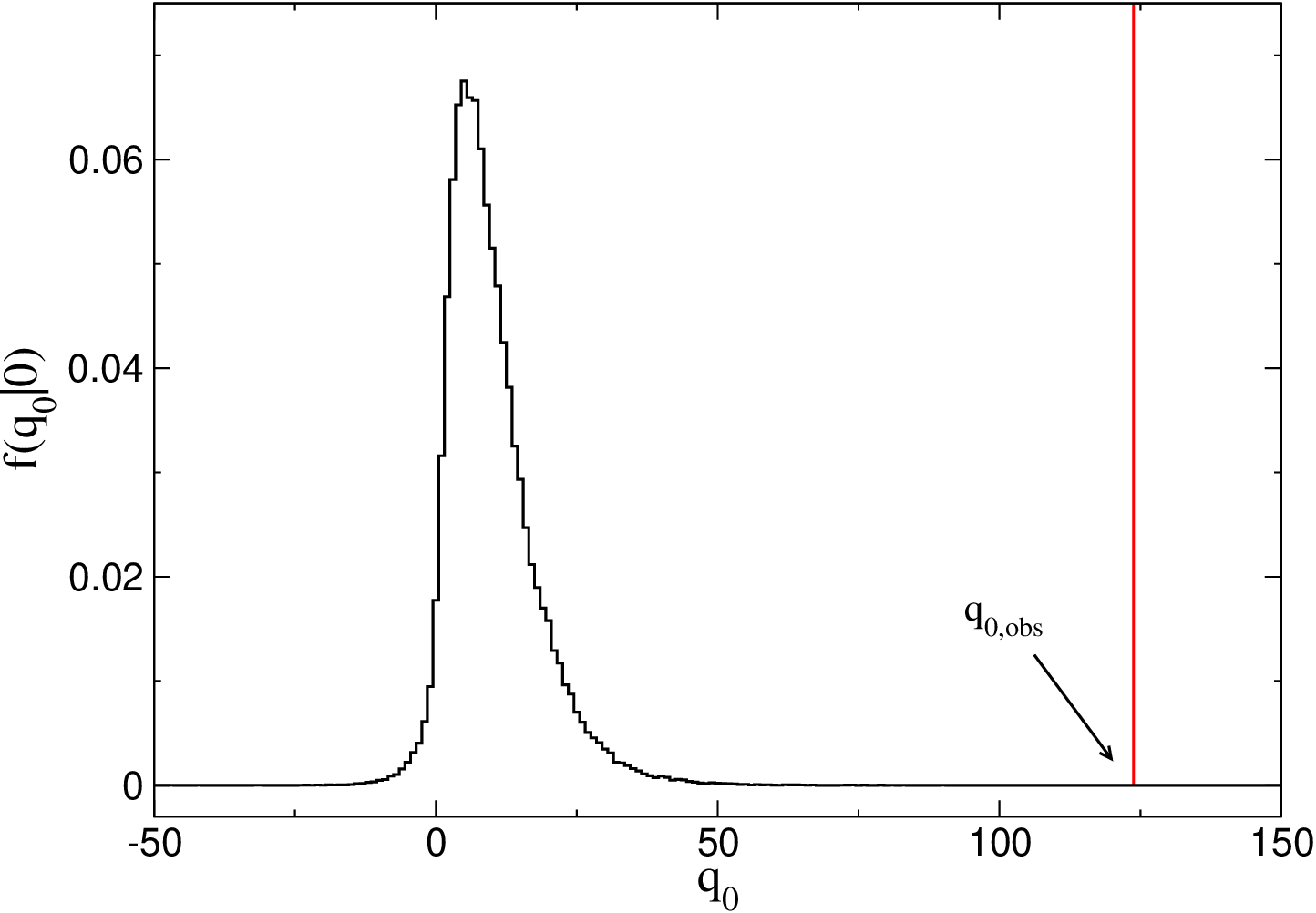}% Here is how to import EPS art
\caption{\label{fig:toy_mc}Histogrammed distribution of $f(q_0|0)$ obtained from $10^5$ synthetic data sets, under the assumption of the background-only hypothesis $(H_0)$ for the $55^\circ$ data. The experimental value is $q_{0,{\rm obs}} = 123.8$.
}
\end{figure}

Complementary to the above approach, one can alternatively (and more robustly) quantify a discovery significance for the 6401~keV state, using the test statistic $q_0 = -2 \ln \lambda(0)$~\cite{Cowan:11}, where $\lambda(0)$ is the profile likelihood ratio (LR)
\begin{equation}
\label{eq:LR}
\lambda(0) = \dfrac{\mathcal{L}(0,\hat{\hat{\bm{\theta}}})}{\mathcal{L}(\hat{\eta},\hat{\bm{\theta}})}. 
\end{equation}
In the above, $\eta$ represents the signal strength of the 6401~keV peak, and is characterized by its area. $\mathcal{L}(0,\hat{\hat{\bm{\theta}}})$ is the profile likelihood~\cite{pdg,Cowan:11} evaluated with $\hat{\hat{\bm{\theta}}}$, the values of $\bm{\theta}$ that maximize the likelihood $\mathcal{L}$, for a specified $\eta = 0$. The denominator is the best-fit likelihood function under $H_1$, with $\eta > 0$. 
The advantage of such a LR test is that it provides the highest power\footnote{The power of a test relates to the probability of rejecting the null-hypothesis when it is false.}~\cite{neyman,pdg} test of $H_0$ with respect of the alternative $H_1$, for a given significance level $\alpha$. Additionally, the ratio also nullifies any systematic effect contributions from assumed peak lineshape parameters, etc.
%\footnote{In fact, our conclusions remain unchanged if we use different values for the tail parameters $l$ and $\xi$, or assume pure Gaussians for the peak shapes.}. 
%
\begin{figure}[t]
\includegraphics[scale=0.34]{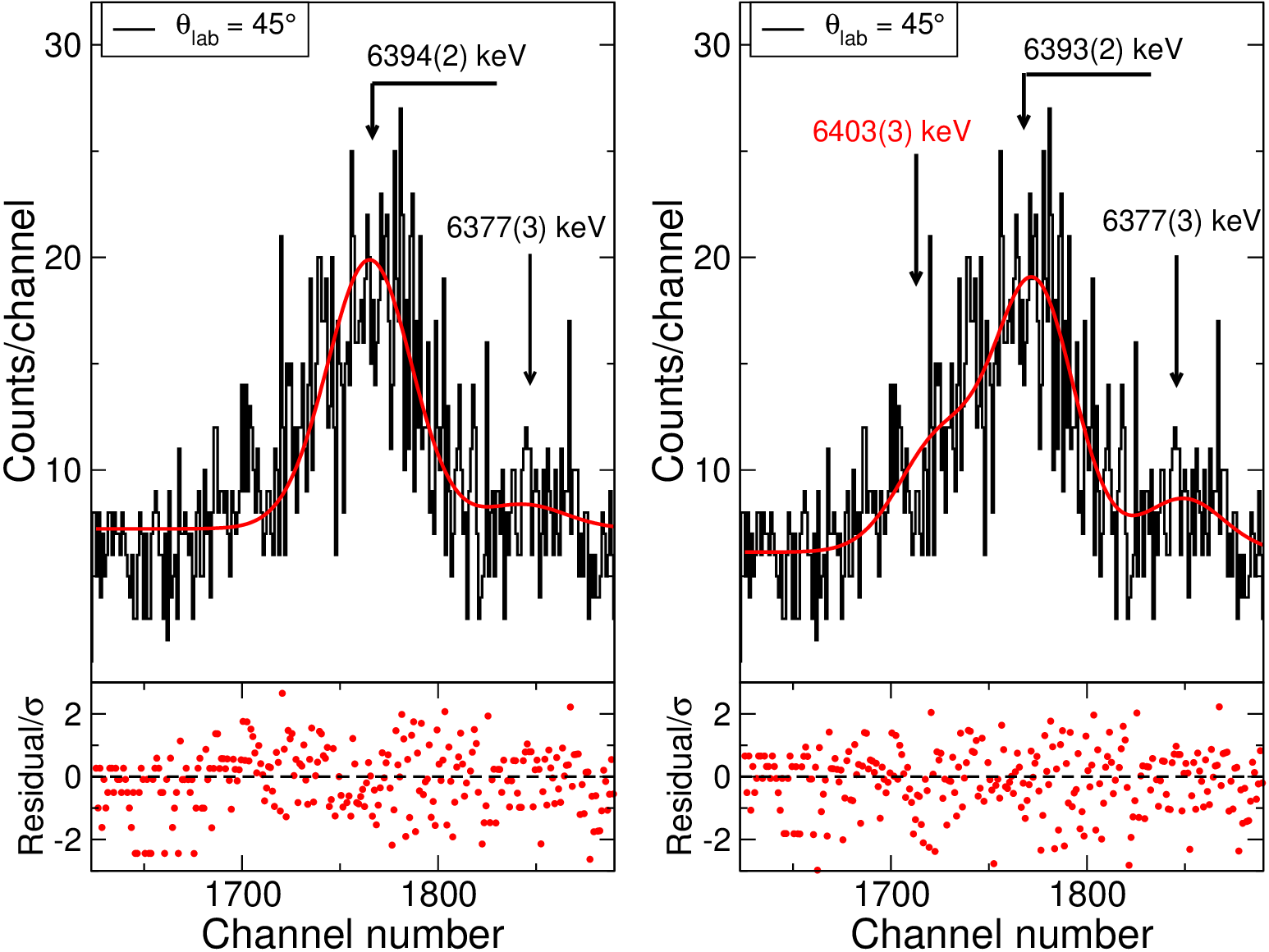}% Here is how to import EPS art
\caption{\label{fig:restricted2}Fits to the $45^\circ$ data, compared similarly as in Fig.~\ref{fig:restricted}.
}
\end{figure} 
% 
% Compared to the above approach, a more rigorous~\cite{neyman} hypothesis test uses the
% %statistic $-2 \ln \tilde{\lambda}(\bm{\theta})$, where $\tilde{\lambda}(\bm{\theta})$ is the 
% likelihood ratio $L(H_0)/L(H_1)$~\cite{Cowan:11,Wu} evaluated for the ML estimates. If the counts $n_i$ in each bin are independent and Poisson distributed, the log-likelihood function is 
% % \begin{equation}
% % L(\bm{\theta}) = \prod_{i=1}^{N~{\rm bins}}\frac{y_i(\bm{\theta})^{n_i}}{n_i!} e^{-y_i(\bm{\theta})}. 
% % \end{equation}
% \begin{equation}
% \ln L(\bm{\theta}) = \sum_{i=1}^{N~{\rm bins}}n_i \ln y_i(\bm{\theta}) - y_i(\bm{\theta}) - \ln n_i{\displaystyle !\,}.
% \end{equation}
% In terms of these log-likelihoods, the likelihood-ratio test statistic is simply
% 
% \begin{equation}
% t_{\bm{\theta}} = -2\left[\ln L(H_0) - \ln L(H_1)\right]. 
% \end{equation}
% %

%
\begin{figure}[t]
\includegraphics[scale=0.34]{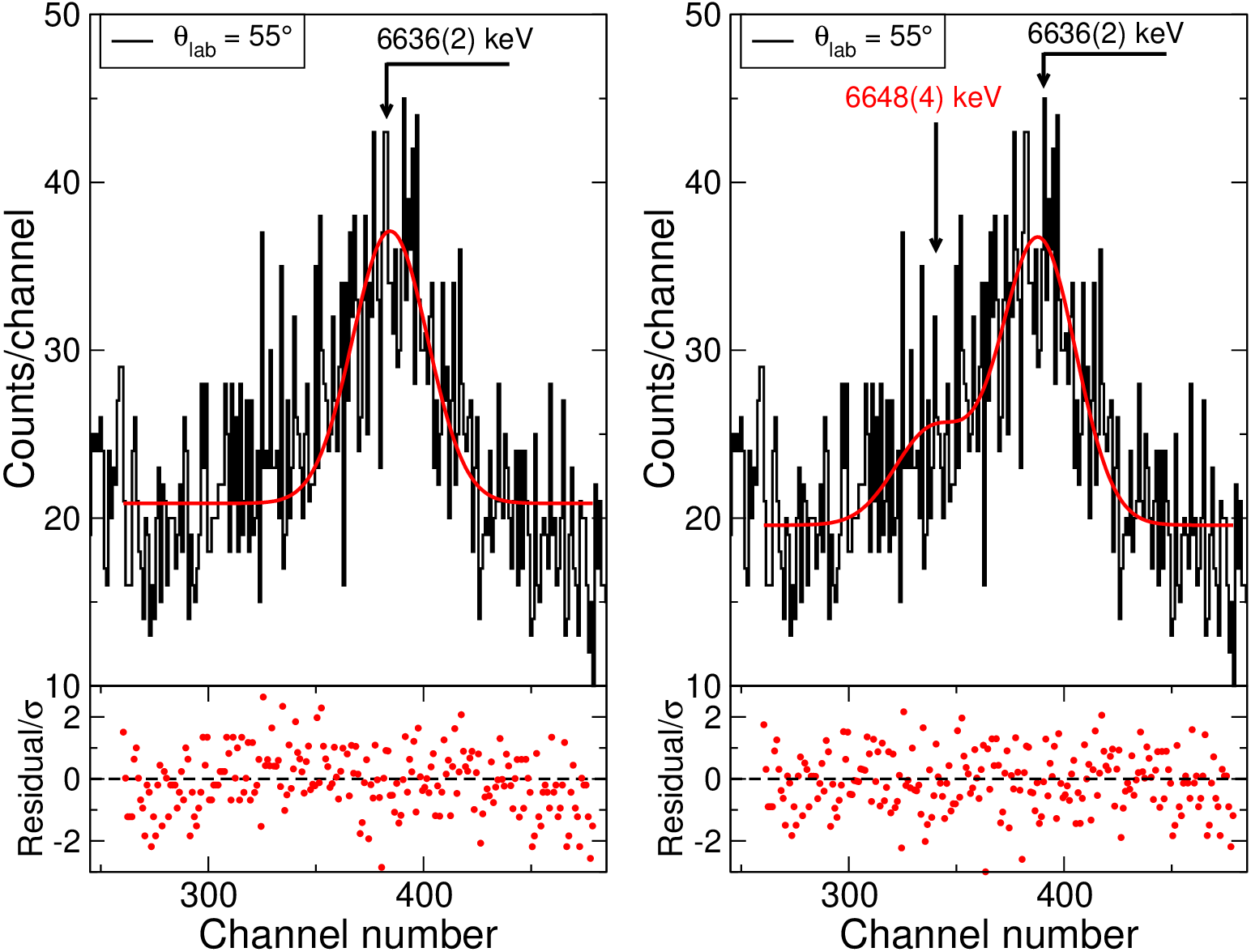}% Here is how to import EPS art
\caption{\label{fig:fits3}Fit results in the 6636~keV region. Left panel: Assuming a single peak corresponding to the 6636~keV state (the null hypothesis). Right panel: On including an additional peak  corresponding to $E_x \approx 6648$~keV (the alternative hypothesis). 
}
\end{figure} 

The $p$-value associated with the test-statistic $q_0$ is simply~\cite{Cowan:11} 
\begin{equation}
p_0 =  \int_{q_0,\mathrm{obs}}^\infty f(q_0|0) dq_0,   
\end{equation}
where $f(q_0|0)$ is the pdf that describes $q_0$, under the assumption of the background-only ($H_0: \eta = 0$) hypothesis. Here again, considering the limited applicability of Wilks' theorem for likelihood ratio tests~\cite{Algeri}, we 
fall back on Monte Carlo simulations to determine the distribution of $f(q_0|0)$. Fig.~\ref{fig:toy_mc} compares the $f(q_0|0)$ values obtained from $10^5$ toy Monte Carlo data sets, with the measured value of $q_0$ for $\theta_{\rm lab} = 55^\circ$. These results yield a $p$-value $< 5 \times 10^{-6}$, which rules out the null hypothesis at the $> 4 \sigma$ level. A similar analysis for the low-statistics data at $\theta_\mathrm{lab} = 45^\circ$ also yields a small probability $p_0 = 0.026$.  We show the fits to these data in Fig.~\ref{fig:restricted2} for completeness, despite the meager statistics at this angle.

In addition to the above, we note that our higher statistics $55^\circ$ spectrum shows an excess of counts in the low energy region of the triton peak corresponding to the 6636~keV state (c.f. Fig.~\ref{fig:spectrum}). Close inspection of the spectra from the previous $^{31}{\rm P}({^3{\rm He},t})$ experiment by Parikh~{\it et al.} (Fig.~1 in Ref.~\cite{Parikh:11}) and the $^{32}{\rm S}(d,t)$ work by Irvine~{\it et al.}~(Fig.~1 in Ref.~\cite{Irvine}) shows some evidence of similar structure. We thus analyzed the data in this ROI similarly as with the 6400~keV region. Available data from Refs.~\cite{Rebeiro1,Rebeiro2,Rebeiro3,Jespere,Kamil} clearly indicated that the peaks in this region ought to have nearly Gaussian lineshapes. 
Using the same approach as before, we first fitted the 6636~keV peak with its width as a free parameter. This yielded an anomalously large FWHM of 65(6) channels, which is nearly twice the expected value of 33(4) channels, and about 1.5 times larger than the 6255~keV peak. 
This disagreement prompted us to proceed as previously, incorporating the extended maximum likelihood method in Eq.~\eqref{eq:min2}, again with conservative $\pm 95\%$ CL constraints on the FWHM parameter. The null hypothesis fit results, together with a new set of toy Monte Carlo simulations for these data, yielded $p_0 = 0.018$. This small $p$-value strongly suggests an additional state in the region, shown in the right panel of Fig.~\ref{fig:fits3}. On including this extra peak in our analysis, we determine the excitation energy of the state to be 6648(4)~keV. This corresponds to a resonance at $E_r = 517(4)$~keV, towards the higher edge of the Gamow window for the $^{30}{\rm P}(p,\gamma)$ reaction at peak nova temperatures. Similarly significant results (that implied new features in the peak structures) were not obtained for any of the other observed triton groups\footnote{We also do not see any signature of the $J^\pi = 3/2^+$, $T = 3/2$ isobaric analog state, which is known to have considerable isospin mixing with the $T = 1/2$ state at 6390~keV~\cite{Bennett:16}. This may be because we were limited by experimental sensitivity.}  in Fig.~\ref{fig:spectrum}. 

\section{Astrophysical implications}

The contribution of $^{30}{\rm P}(p,\gamma)$ resonances can be  determined from their individual resonance strengths  
\begin{equation}
\omega \gamma = \frac{(2J_r+1)}{6}\frac{\Gamma_p \Gamma_\gamma}{\Gamma},
\end{equation}
where $J_r$ is the spin of the resonance, $\Gamma_p$ and $\Gamma_\gamma$ are the partial proton and gamma widths of the resonant state in $^{31}$S, and $\Gamma = \Gamma_p + \Gamma_\gamma$ is its total width. Although past experimental investigations~\cite{Bennett:16,Ma:07,Wrede:07,Wrede:09,Parikh:11,Irvine,Anu1,Bennett:18,Anu2,Doherty1,Doherty2,Setoodehnia,Jenkins} have provided invaluable information in this regard, the spins and parities of only three of these levels are known with certainty at present. These include~\cite{ensdf} the $J^\pi = 3/2^+,~T = 3/2$ isobaric analog state (IAS) at 6281~keV, the $1/2^+$ state at 6255~keV and a recently reported $3/2^+$ state at 6390~keV~\cite{Bennett:16}. The latter was  identified via accurate measurements of $\beta$-delayed $\gamma$ ray intensities, together with shell model calculations that showed significant isospin mixing of the ($T = 1/2$) state with the $T = 3/2$ IAS. This 6390~keV state (at $E_r = 259$~keV) is arguably the most important $^{30}{\rm P}(p,\gamma)$ resonance~\cite{Bennett:16} known at the present time, with a calculated  resonance strength of 24~$\mu$eV~\cite{Bennett:16}. Very recently a revised $\omega \gamma = 80(48)~\mu$eV~\cite{Budner} was reported for this resonance, from a measurement of the proton branching ratio for the state. 
\begin{figure}[t]
\includegraphics[scale=0.34]{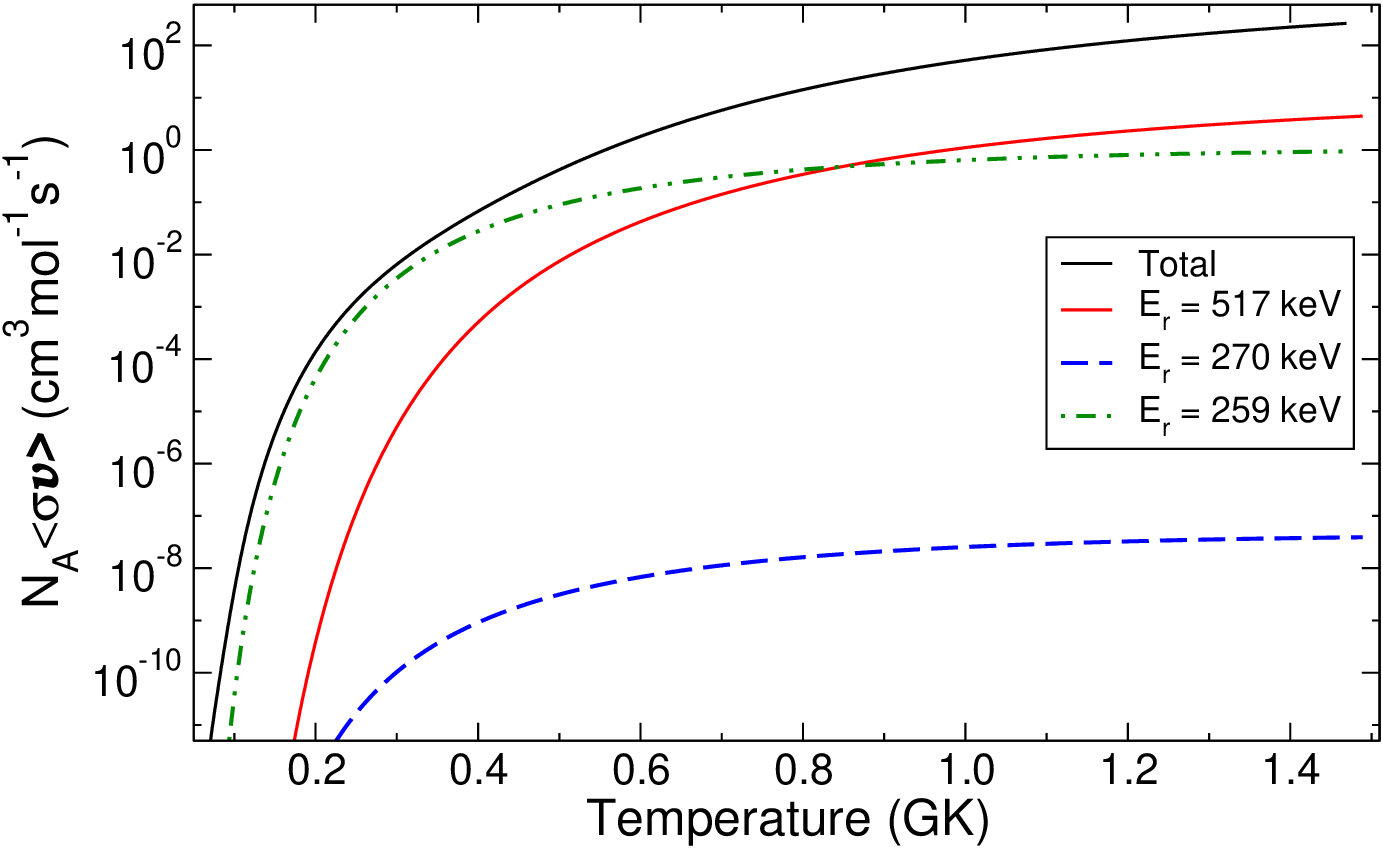}
% Here is how to import EPS art
\caption{\label{fig:rates} Calculated $^{30}{\rm P}(p,\gamma)$ reaction rates for the two resonances discussed in this work, at $E_r = 270$~keV  and $E_r = 517$~keV. For comparison we also show the contribution of the important 259~keV resonance reported in Refs.~\cite{Bennett:16,Budner}. Although the $\omega\gamma$ for this resonance is reported with 60\% relative uncertainty~\cite{Budner}, we evaluate the rate based on its central value. We forgo uncertainty bands in this plot because of the reasons specified in Ref.~\cite{Iliadis}.  The total reaction rate is for all significant resonances, obtained using the shell model results from Refs.~\cite{sm1,sm2} when experimental information was lacking. 
}
\end{figure} 

Because of the limited available experimental information regarding most the proton unbound states in $^{31}$S, the authors of Refs.~\cite{sm1,sm2} performed comprehensive shell model calculations using the USDB-cdpn Hamiltonian, that included a description of several negative parity states. They matched around 20 theoretically predicted levels (in the 5.9--7~MeV range) with experiment, and calculated the $\omega\gamma$ for each of the resonances to obtain the reaction rate over $0.1 \le T_9 \le 10$. In their analysis, the 6401~keV resonance was matched to a $7/2^+$ state predicted at 6298~keV, with calculated $\Gamma_p = 2.7\times10^{-15}$~keV and $\Gamma_\gamma = 5.3\times10^{-5}$~keV. However, two $1/2^-$ states predicted at 6247 and 6602~keV could not be associated with any experimentally levels reported during the time. 
%Their resonance strengths were calculated to be $1.0 \times 10^{-13}$ and $2.9 \times 10^{-6}$~keV respectively. 
It is therefore quite likely that the latter level is the 6648(4)~keV 
state identified in this work. Its resonance strength was calculated to be  $\omega \gamma = 2.9 \times 10^{-6}$~keV.

Fig.~\ref{fig:rates} compares the individual contributions of the 270 and 517~keV resonances to the recently reported 259~keV resonance~\cite{Bennett:16,Bennett:18} that is expected to significantly contribute to the total reaction rate over $0.1 \le T_9 \le 0.4$. We note that with the recent shell-model-evaluated resonance strength, the contribution of the 6401~keV state ($E_r \approx 270$~keV) is much smaller than the other two resonances over a large range of temperatures. In comparison, our matched $1/2^-$ 6648~keV level has a relatively larger contribution, beginning to get influential around $T_9 \approx 0.6$. This may have significant repercussions.  
While most nova models show maximum peak temperatures of about 0.4~GK, it has been shown that cold massive white dwarfs ($M > 1.2 M_{\odot}$) with low accretion rates~\cite{Townsley_2004} can lead to peak nova temperatures of around 0.5~GK~\cite{Glasner_2009}. In such cases the contribution of this resonance cannot be neglected. Furthermore, the $^{30}{\rm P}(p,\gamma)$ reaction rate is also an important ingredient in the modeling of X-ray bursts~\cite{xrb}, where the peak temperatures approach around 1.4~GK. At these temperatures the fractional contribution
of our proposed 517~keV resonance to the total reaction rate is much higher. 

\section{Summary}
We used a $^{32}{\rm S}(d,t)$ measurement to study the excitation energy region in $^{31}$S relevant for the $^{30}{\rm P}(p,\gamma)^{31}{\rm S}$ reaction in classical novae. A likelihood ratio analysis of our data supports the existence of a 6401~keV state in $^{31}$S, claimed to be observed in previous $^{32}{\rm S}(d,t)$ and $^{31}{\rm P}(^3{\rm He},t)$ work. The data also suggest, with high significance, an additional state in $^{31}$S at about 6648~keV. This corresponds to a $^{30}{\rm P}(p,\gamma)$ resonance at around 517~keV, towards the higher edge of the Gamow window at peak nova temperatures of $\sim$~0.4~GK. 

In light of these results, further investigations to corroborate the existence of 6648~keV state in $^{31}$S and determine its properties (spin-parity, lifetime, proton width, etc.) would be welcome. Similar work related to the 6401~keV state also remain well motivated. 

\begin{acknowledgments}
We thank Alex Brown for the sharing with us the complete shell model results from Refs.~\cite{sm1,sm2}. Useful discussions with Chris Koen, Alejandro Garc\'ia and Werner Richter are gratefully acknowledged. This work was partially funded by the National Research Foundation (NRF), South Africa under Grant No. 85100 and the Natural Sciences and Engineering Research Council (NSERC) Canada.   
\end{acknowledgments}

\bibliography{32Sdt_resubmit_april}% Produces the bibliography via BibTeX.

\end{document}